\documentclass[twocolumn,prl,10pt,aps,longbibliography,superscriptaddress]{revtex4-2}

\usepackage{amsmath}
\usepackage{amssymb}
\usepackage{amsthm}
\usepackage{dsfont}
\usepackage{mathtools}
\usepackage{graphicx} 
\usepackage{parskip}
\usepackage{appendix}
\usepackage{hyperref}
\usepackage{xcolor}

\newcommand{\red}[1]{{\color{red}#1}} 

\newcommand{\rs}{_\text{rest}}

\usepackage{xstring}
\newcommand{\refcite}[1]{
\begingroup
\def\tempx{0}
  \StrCount{#1}{,}[\tempx]
  \ifnum\tempx > 0 
  Refs.~%
  \else
  Ref.~%
  \fi
\endgroup
\cite{#1}
}

\newcommand{\rvline}{\hspace*{-\arraycolsep}\vline\hspace*{-\arraycolsep}}

\begin{document}

\title{Topological Properties of Single-Particle States\\ Decaying into a Continuum due to 
Interaction}

\author{B. Hawashin}
 \email{bilal.hawashin@tu-dortmund.de}
 \affiliation{
 Condensed Matter Theory, Department of Physics, TU Dortmund University, 
Otto-Hahn-Stra\ss{}e 4, 44221 Dortmund, Germany}
 \affiliation{
 Department of Physics and Astronomy and Manitoba Quantum Institute, 
University of Manitoba, Winnipeg R3T 2N2, Canada}
\author{J. Sirker}
 \email{sirker@physics.umanitoba.ca}
 \affiliation{
 Department of Physics and Astronomy  and Manitoba Quantum Institute, 
University of Manitoba, Winnipeg R3T 2N2, Canada}
\author{G. S. Uhrig}
    \email{goetz.uhrig@tu-dortmund.de}
    \affiliation{
 Condensed Matter Theory, Department of Physics, TU Dortmund University, 
Otto-Hahn-Stra\ss{}e 4, 44221 Dortmund, Germany}

\begin{abstract}
    We investigate how topological Chern numbers can be defined when single-particle states 
		hybridize with continua. We do so exemplarily in a bosonic Haldane model 
		at zero temperature with an additional on-site decay of one boson into two and the conjugate fusion 
		of two bosons into one. Restricting the Hilbert space to two bosons at maximum, 
		the exact self-energy is accessible. We use the bilinear Hamiltonian $H_0$ corrected by the self-energy 
		$\Sigma$ to compute Chern numbers by two different approaches. 
		The results are gauged against a full many-body calculation in the  Hilbert space
		where possible. 
		We establish numerically and analytically that the effective Hamiltonian 
		$H_\text{eff}=H_0(\vec k) +\Sigma(\omega,\vec k)$ reproduces the correct many-body topology 
		if the considered band	does not overlap with the continuum. In case of overlaps, one can extend the definition of the
		Chern number to the non-Hermitian $H_\text{eff}$ and there is evidence that the Chern number
		changes at exceptional points. But the bulk-boundary correspondence appears to be no longer valid
		and edge modes delocalize.
\end{abstract}
 
\date{\today}

\maketitle


\paragraph{\textbf{Introduction}}
The first striking topological effect discovered in condensed matter physics 
is the integer quantum Hall effect \cite{klitz80}. Here, the number of contributing
edge modes defines the Hall conductivity \cite{thoul82,kohmo85} to an incredible precision 
revolutionizing metrology \cite{weis11}.
No non-linearities occur \cite{klein90,uhrig91}. Also, fractional charges are possible
enabling the fractional quantum Hall effect \cite{tsui82}. Both topological effects
gained their discoverers Nobel Prizes \cite{nobel1985,nobel1998} and topology in 
condensed matter has continued to gain great interest in the last decades \cite{nobel2016}. 
The possible types of topological order have been classified for Gaussian fermionic systems, 
both in the case of non-spatial and spatial symmetries 
\cite{SchnyderRyuPRB,RyuSchnyder,RyuSchnyderReview,Bernevig2011,Bernevig2013,Bernevig2014}. 
A non-trivial topological ground state of a fermionic insulator cannot be connected adiabatically 
to the atomic limit without closing the gap or, if present, breaking the symmetry protecting the 
topological order. These topological states thus have non-trivial entanglement and typically 
also robust edge states \cite{RyuHatsugaiChiralEntanglement,Fidkowski,MonkmanSirker1,MonkmanSirker3,MonkmanSirker4}. 
For quantum Hall systems, in particular, topologically non-trivial ground states were 
proposed even for models without external magnetic field \cite{halda88b} and realized 
in experiment \cite{konig07,ando13,jotzu14,yue19}. Numerous technological applications 
are expected not only in fermionic systems \cite{kong10,kong11,tokur19,frolo20} but also 
in insulating quantum magnets, see e.g.\ Ref.\ \cite{malki20b}.

A crucial aspect of quantum Hall systems breaking time reversal symmetry is the occurrence 
of robust, chiral edge modes which generically allow for motion
only in one direction. These systems promise a high degree of tunability \cite{malki17b,wang18b}.
The number of these chiral edge modes is linked
to the Chern number of a dispersive band, i.e., a topological invariant given by the Berry phase 
of a quantum state tracked around the Brillouin zone (BZ). This is commonly referred to as the bulk-boundary correspondence \cite{hasan10,berne13}. However, already for non-interacting
systems, this correspondence must be used
with caution because it only holds if the dispersive band is protected by \emph{indirect}
energy gaps \cite{malki19c,malki19b}. Otherwise, the potential boundary mode is not
localized, but extends into the continuum. Such a breakdown of the bulk-boundary
correspondence can also occur in manifestly non-Hermitian systems \cite{shen18,bergh21,wang22,okuma23,MonkmanSirkerNH}.

Thus, the hybridization of modes
with continua can destroy the desired topological properties. Clearly, if a single-particle
excitation is no longer defined in the whole BZ, its Berry connection \cite{berry84}
cannot be defined in the standard way. Such a hybridization of single-particle states with 
continua is generic in interacting systems. Hence, a generalization of topological invariants
is required. For ground states of fermionic topological insulators, 
topological invariants such as the Chern number can be expressed in terms of 
Green's functions at zero frequency \cite{niu85,wang10d,gurar11}.
Clearly, this allows for a straightforward extension to systems with interactions although
zeroes in the Green's function need to be considered carefully. The appearance of
edge modes at zero energy at the boundary between two topologically distinct phases
is then the generic scenario, supporting the bulk-boundary correspondence. Similarly, a classification of symmetry-protected topological phases in interacting bosonic systems based on group cohomology theory exists \cite{doi:10.1126/science.1227224}.

Here, we do not focus on the ground state and its topological properties but rather on 
the bands of elementary, single-particle excitations. Without interactions, the Berry curvature 
is unambiguously defined. Including interactions, two situations need to be
distinguished: (i) If the single-particle states are adiabatically connected
to the ones in the absence of interactions, i.e, the hybridization renormalizes
the single-particle states, then the Chern number can be defined in a mathematically rigorous way
and the bulk-boundary correspondence holds if the edge mode is protected
by an energy gap. (ii)
If the single-particle states hybridize strongly with the continua, so that they merge
with them, a Chern number can be defined for the non-Hermitian effective Hamiltonian, but
not for the full Hamiltonian. The bulk-boundary correspondence appears not to be valid in this case. 

We define a bosonic model reduced to the essentials which allows for various approaches
to treat it. Our study provides evidence that the self-consistent solution
of the effective single-particle problem defined by the non-interacting
Hamiltonian plus the proper self-energy is the appropriate generalization to calculate Chern numbers in the interacting case.
This approach provides a connection between many-particle Hermitian models and 
single-particle non-Hermitian models.
We emphasize, however, that usually non-Hermiticity is invoked on the basis of 
gains and losses. Here, in contrast, we investigate a Hermitian many-body
problem which can be reduced to an effective single-particle problem if the
complex self-energy with its full frequency dependence is known. This effective problem 
can be non-Hermitian, but its non-Hermiticity depends on frequency. 

\paragraph{\textbf{Model}} We consider interacting bosons on a 
honeycomb lattice with sublattices $A$ and $B$ sites described by 
\begin{subequations}
\label{eq:modeldef}
\begin{align}
       H &=  {H}_0 +  {H}_\text{int} \\
     {H}_0 &= t_1 \sum_{\langle i,j \rangle} \left(  {b}_i^\dagger  {b}_j + \text{h.c.} \right) + t_2 \sum_{\langle \langle i,j \rangle \rangle} \left( e^{i \nu_{ij} \phi}  {b}_i^\dagger  {b}_i + \text{h.c.} \right) \nonumber \\
    & \quad + M \sum_i \varepsilon_i  {b}_i^\dagger  {b}_i + E_0 \sum_i  {b}_i^\dagger  {b}_i 
		\\
      {H}_\text{int} &= g \sum_i \left(  {b}_i^\dagger  {b}_i^\dagger  {b}_i + \text{h.c.} \right),
\end{align}
\end{subequations}
where $ {b}_i$ and $ {b}_i^\dagger$ are bosonic annihilation and creation operators, respectively.
The prefactors $t_1, t_2, g, E_0 \in \mathbb{R}^+$ and $M \in \mathbb{R}$ are energies, $\nu_{ij} = + 1$ if hopping from $i$ to $j$ is clockwise in a hexagon and $\nu_{ij} = - 1$ otherwise. The local
sign is $\varepsilon_i = 1$ if $i \in A$ and 
$\varepsilon_i = -1$ if $i \in B$. The sums over $\langle i,j \rangle$ and 
$\langle \langle i,j \rangle \rangle$ run over nearest neighbors and next-nearest neighbors, respectively.  This bilinear part is analogous to the fermionic
Haldane model which displays well-established  topological properties with finite
Chern numbers \cite{halda88b}. The on-site energy $E_0$ \red{serves} two purposes: first, $E_0$
needs to be sufficiently large so that the bosonic excitation energies are positive. Then,
the ground state is the topologically trivial bosonic vacuum.
Our focus is on the topological properties of the excited bands.
Second, tuning $E_0$ and $g$ allows to  control the strength of 
the hybridization between one-boson and two-boson states.

The cubic interaction $ {H}_\text{int}$ describes the decay of one boson into two bosons on the same site with coupling constant $g$ or the inverse process, i.e., the fusion of two bosons into one.
Terms of this kind occur generically in quantum magnets for spin waves in case of non-collinear order
\cite{chern09,chern16,zhito13} as well as for triplons in valence bond solids 
\cite{zhito06,fisch11,zhito13,mulle23a}. We stress that one should first perform
a Bogoliubov transformation to diagonalize the bilinear part including terms of two annihilation or 
creation operators. Otherwise, it is unavoidable to generalize the operator scalar product to
a symplectic or paraunitary one, see for instance Refs.\ \cite{shind13b,peano16,malki19c}.  
The diagonalized Hamiltonian is
similar to \eqref{eq:modeldef} except that the hopping terms and the interaction terms
have a certain spatial range. In this sense, $ H$ is an idealization adopted here not
to be distracted by a plethora of parameters.
We consider the cubic interaction
because the continuum to which the single-particle states
are coupled  is then governed by only one free momentum while
the relevant continuum for the quartic interaction requires two free momenta.
Thus, the treatment of cubic terms is considerably easier than the one
of quartic terms.
The possibility of having cubic interactions is also a reason to consider a bosonic 
instead of a fermionic model where cubic terms do not occur in standard systems.

In order to deal with an exactly solvable limit, we restrict the Hilbert space to 
$\mathcal{H} = \mathcal{H}(0) \oplus \mathcal{H}(1) \oplus \mathcal{H}(2)$,
where $\mathcal{H}(n)$ is the Hilbert space of exactly $n$ bosons.
Hence, the restriction allows for two bosons at maximum. 
Then, we can exactly determine the self-energy at zero temperature,
 see the Suppl.~Mat.~\cite{supplement}. The 
efficient evaluation on meshes in the BZ of moderate sizes is carried out
by means of a generalized continued fraction expansion derived in the Suppl.~Mat.~\cite{supplement}.
Alternatively, we can consider the self-energy as the leading contribution in perturbation theory of
quadratic order in $g$. In this view, the Hilbert space does not need to be restricted.

\paragraph{\textbf{Chern numbers}}
For non-interacting systems, the Chern number of the $n$-th band is defined as 
$C = \frac{1}{2\pi} \int_\text{BZ} F_{12} d^2k$, where 
$F_{12} = \partial_1 A_2 - \partial_2 A_1$ denotes the Berry curvature and 
$A_i(\vec{k}) = \langle u_n (\vec{k}) | \partial_i |u_n(\vec{k}) \rangle$ is the Berry connection. 
The states $|u_n(\vec{k}) \rangle$ are the single-particle states at wave vector $\vec{k}$.
In order to calculate these states, the coefficient matrix $H_0(\vec k)$ of the bilinear Hamiltonian
at this wave vector is sufficient. For the model under study, this amounts to the solution
of a $2\times2$ problem at each given point of a mesh in the BZ.
In our calculations, we use meshes ranging from $20\times20$ to $30\times30$.
These sizes are sufficient to compute the Chern numbers \cite{fukui05}.

We proceed in three distinct ways:
\paragraph{\textbf{a) Topological approach}}
    We use the eigenstates of 
		\begin{subequations}
		\begin{equation}
		H_\text{top}(\vec{k})= H_0(\vec k)+\Sigma(\omega=0,\vec k)
		\end{equation}
		fulfilling 
    \begin{equation}
		\label{eq:top}
        H_\text{top}(\vec{k}) |u_n(\vec{k})\rangle = E_n(\vec{k}) |u_n(\vec{k})\rangle.
    \end{equation}
		\end{subequations}
		Obviously, this is still a $2\times2$ problem. At zero frequency, the
		self-energy is Hermitian so that no issues from non-Hermiticity arise; $E_n(\vec{k})$
		is real.	Wang et al.\ showed that for fermionic models at $T = 0$
		the sum $\sum_{\alpha \text{ occ.}} C_\alpha$
		of the Chern numbers $C_\alpha$ obtained from $H_\text{top}(\vec{k})$ for the occupied bands
		provides the Chern number of the ground state which defines the Hall conductivity \cite{wang10d,wang12}.
\paragraph{\textbf{b) Effective approach}}  
    We use the eigenstates of 
		\begin{subequations}
		\begin{equation}
		H_\text{eff}(\omega,\vec{k})= H_0(\vec k)+\Sigma(\omega,\vec k)
		\end{equation}
		fulfilling the self-consistent eigenvalue equation
    \begin{equation}
		\label{eq:eff}
        H_\text{eff}(E,\vec{k}) |u(E, \vec{k})\rangle = E(\vec{k}) |u(E,\vec{k})\rangle.
    \end{equation}
		\end{subequations}
		This is still a $2\times2$ problem, but with the calculation of the self-consistent $E(\vec{k})$ 
		as additional challenge. For non-Hermitian diagonalizations, the right and left eigenstates differ,
		but they define fiber bundles over the manifold given
by the torus of the BZ, allowing for the definition of quantized Chern numbers 
		and it has been shown that they yield the same Chern numbers \cite{shen18}. 
		In Ref.\ \cite{wang13c}, the suitability of the 
		effective Hamiltonian $H_\text{eff}(\omega, \vec{k})$ was critically discussed 
		because it did not provide the correct Chern number of fermionic ground states.
		Since, however, our focus is a different one it is worthwhile considering $H_\text{eff}(E,\vec{k})$.
\paragraph{\textbf{c) Two-body approach}}
    We use the eigenstates at a given total wave vector $\vec{k}$
		in the total Hilbert space $\mathcal{H}$
		\begin{equation}
		\label{eq:2body}
          H(\vec{k}) |u_n(\vec{k})\rangle = E_n(\vec{k}) |u_n(\vec{k})\rangle.
    \end{equation}
		This is a standard Hermitian diagonalization problem, but in a large Hilbert space
		of which the dimension at fixed wave vector $\vec{k}$ is given by the number of sites in the model. Numerically, we 
		deal with sizes of 100 to 1000 sites. The calculation of Berry
		connections does not pose a conceptual difficulty in the total Hilbert space
		as long as one can uniquely identify the relevant eigenstates. This is the crucial obstacle
		in the case of bands overlapping with the continuum.

\paragraph{\textbf{Phase transition at gap closure}}
In a two-band system, the Chern number of a band does not change if it is only
adiabatically modified while staying separated from the other band by a finite gap. Thus, 
a phase transition between regions of different Chern numbers must be accompanied by the vanishing of the direct energy gap between the two bands, i.e., at some $\vec{k}$
\begin{equation}
\label{eq:gap-close}
    \Delta E(\vec{k}) = |E_1(\vec{k}) - E_2(\vec{k})| = 0.
\end{equation}
In the Haldane model, the gap closes for the chosen parameters $t_1 = t_2 = 1$, $\phi = \pi / 2$ at the 
$K$ point if $M = 3\sqrt{3}$. We vary $M$
keeping the other parameters fixed and
determine $M_C$ where the gap closes. Then, we determine the Chern number
 in the regions $M < M_C$ and $M > M_C$.

\begin{figure}[htb]
    \centering
				\includegraphics[width=\columnwidth]{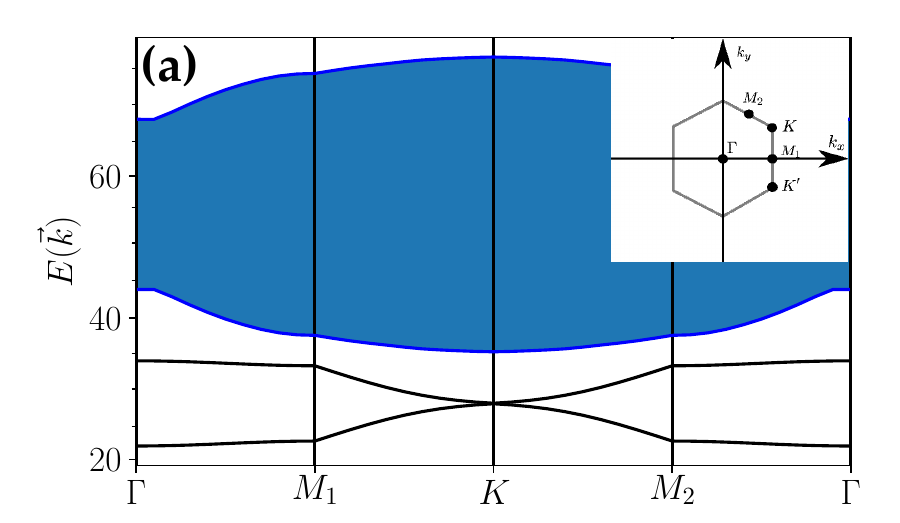}
				\includegraphics[width=\columnwidth]{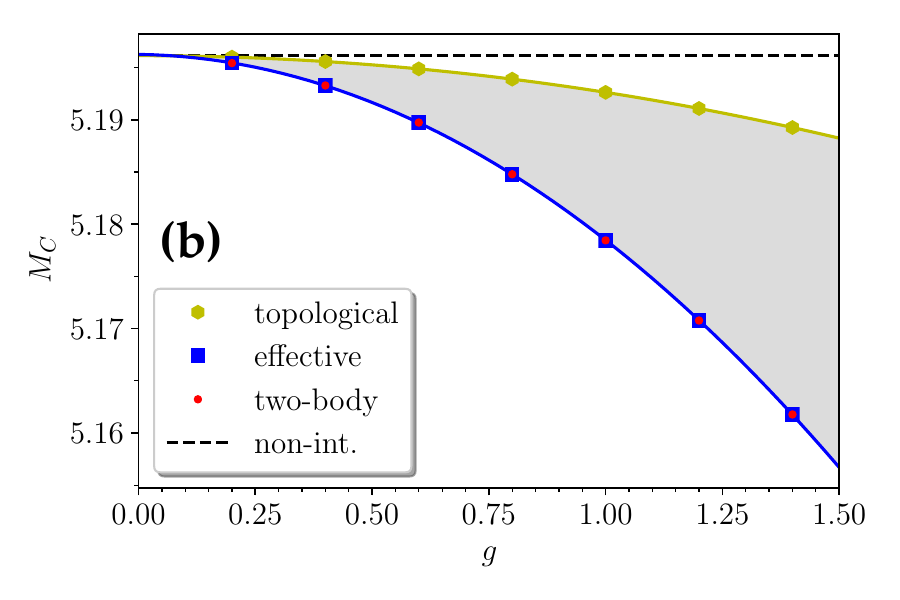}
    \caption{(a) Two-boson continuum (colored) and renormalized single-boson bands 
		for $E_0=28$, $M = 5.2$, and $g=1.4$. 
		(b) Critical values of $M_C$ where the gap between the 
		single-particle bands closes in the three approaches.  Lines are quadratic fits.}
    \label{fig:nool}
\end{figure}

\paragraph{\textbf{Edge states}}
To assess the implications of non-trivial Chern numbers in the interacting case and hence the validity of the bulk-boundary correspondence, 
we calculate the inverse participation ratio (IPR) \cite{krame93,calix15} 
on a honeycomb ribbon periodic in $x$ direction
with length $N_x$ and finite width $N_y$ in $y$ direction with open boundary conditions,
for details see Suppl.~Mat.~\cite{supplement}.
If the $n$-th eigenstate is localized, the IPR stays finite $\lim_{N_y\to \infty} I_n = \text{const}$ while it vanishes $\lim_{N_y\to \infty} I_n \sim 1/N_y^\alpha \to 0$ with $\alpha > 0$ for a delocalized state 
\cite{malki19c} allowing us to determine whether or not a state is localized.
Further evidence for (de) localization 
is obtained from the local density of states (LDOS), also provided in the Suppl.~Mat.~\cite{supplement}.

\paragraph{\textbf{Case I: No overlap with the lower band}}
For large $E_0 = 28$ the bands do not overlap with the two-boson continuum, see Fig.~\ref{fig:nool}(a).
Panel (b) shows the dependence of $M_C$ on the coupling $g$.
The effect is small because the system is in the perturbative regime where we estimate
the self-energy to be $|\Sigma_{ij}| \approx \frac{g^2}{E_0} = 0.07$ for $g = 1.4$ and $E_0 = 28$.
Within numerical accuracy, we find that the gap closure occurs at the $K$ point in the BZ
in all three approaches.
We find $\Sigma(\omega,\vec{k})^\dagger = \Sigma(\omega,\vec{k})$ if 
${\omega < E^{(2)}_\text{min}(\vec{k}) = \underset{\alpha, \beta, \vec{q}}{\text{min}} 
\left( E_\alpha(\vec{k} + \vec{q}) + E_\beta(-\vec{q}) \right)}$. 
Thus, we expect and observe only a (weak) renormalization of the non-interacting bands 
and an infinite lifetime of the excitations. 

Due to the absence of overlap between the one-boson bands and the continuum, 
 the Chern numbers are well-defined in all three approaches.
We find that $C_\text{top} = C_\text{eff} = C_\text{2body} = 1$ below the grey shaded area in 
Fig. \ref{fig:nool}(b), while $C_\text{top} = C_\text{eff} = C_\text{2body} = 0$ above. 
However, the approaches differ within the shaded area 
where $C_\text{top} = 1$ but $C_\text{eff} = C_\text{2body} = 0$. 
The many-body calculation considers the complete quantum state and thus the correct
transition amplitudes enter the Berry connection. Hence, we conclude from the agreement
of the effective approach with the two-body approach that the effective approach is
able to assess the Berry curvature and thus the Chern numbers of the many-body problem 
\emph{without} conducting an extensive calculation in the 
complete Hilbert space. The deviating topological approach, while being justified for
the ground state of fermionic models with filled bands, is not appropriate for
determining the Berry curvature of excitations above the ground state. 

The finding, that the effective Hamiltonian including the self-consistency
yields exactly the same Berry phases and thus Chern numbers as considering
the full Hilbert space, is analytically shown by the following argument.
If the total Hilbert space $\mathcal{H}$ can be split into the direct sum of
a single-particle part $\mathcal{H}(1)$ and the rest $\mathcal{H}\rs$ so that
$\mathcal{H}= \mathcal{H}(1)\oplus\mathcal{H}\rs$ then a generic eigenvector $v$
consists of $v=v_1 + v\rs$. In the Supplement we show that $v_1$ can be found
by solving the self-consistency \eqref{eq:eff}.
 Transporting $v$ parallelly around a contour $\gamma$
yields $v'$ and $\langle v|v'\rangle=\exp(i\varphi)$ implies $v'= \exp(i\varphi) v$
and the Berry phase $\varphi$. Thus, in $\mathcal{H}(1)$ the relation $v_1'=\exp(i\varphi) v_1$ 
holds so that the Berry phase determined in the subspace $\mathcal{H}(1)$
is identical to the one in the total Hilbert space, see also Suppl.~Mat.~\cite{supplement}.

In addition, we checked the bulk-boundary correspondence by inspecting the IPR.
In case (i) we find clear numerical evidence that localized edge modes exist
where the Chern number is different from zero; this is also confirmed by the LDOS, 
see Suppl.~Mat.~\cite{supplement}.

We find the same qualitative behavior if the upper band overlaps with the continuum,
but not the lower band. Then the Chern number of the lower band is still unambiguously
defined and the bulk-boundary correspondence holds.

\begin{figure}[htb]
    \centering
			\includegraphics[width=\columnwidth]{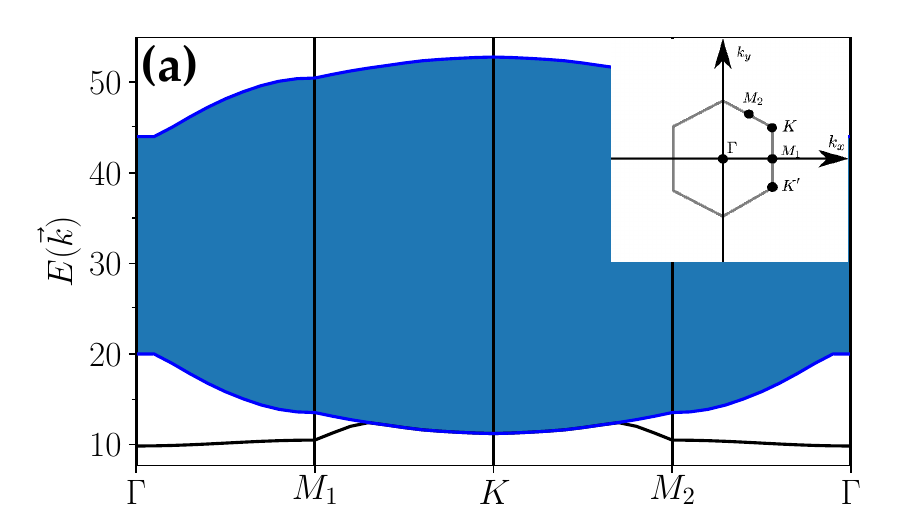}
      \includegraphics[width=\columnwidth]{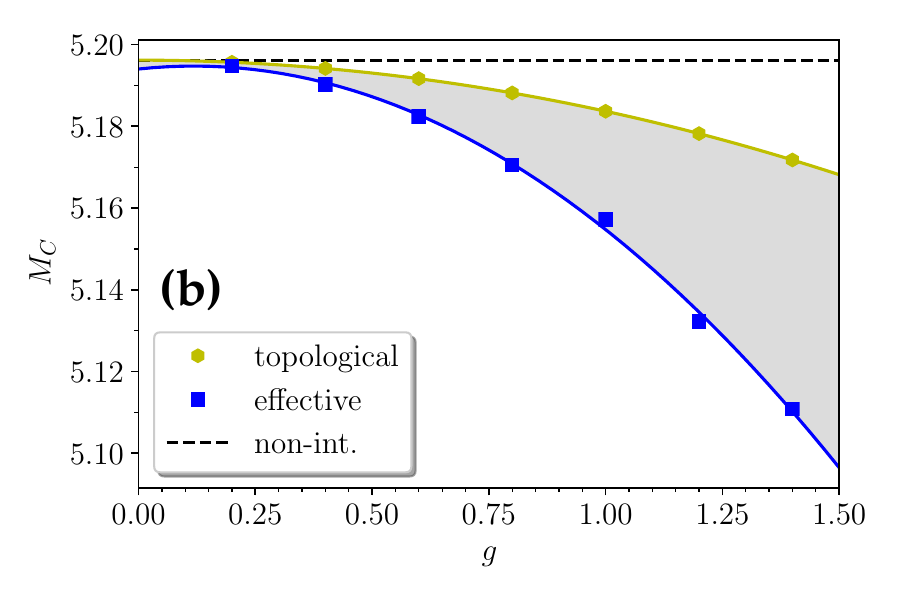}
    \caption{(a) Two-boson continuum (colored) and renormalized single-boson bands 
		for $E_0=16$, $M = 5.2$, and $g=1.4$. (b) Critical values of $M_C$ where the gap closes. 
 Lines are quadratic fits.}
    \label{fig:bool}
\end{figure}

\paragraph{\textbf{Case II: Overlap with both bands}}
If both bands overlap with the continuum as
shown in Fig.\ \ref{fig:bool} for $E_0 = 16$ and $M = 5.2$, it is not a priori clear 
if topological properties survive. The topological approach still works
as before and yields the upper curve $M_C$ in Fig.\ \ref{fig:bool}(b).
But in the two other approaches we encounter major differences to the previous case.
In the effective approach, the self-energy becomes non-Hermitian, but the effective
Hamiltonian still defines two bands. Their eigenvalues are sufficiently separated in the
complex plane so that the eigenvectors are defined unambiguously, except where the
gap closes \eqref{eq:gap-close}, yielding the lower curve in Fig.\ \ref{fig:bool}(b).
We found numerical evidence that at gap closure, i.e., for $E_1(\vec{k}) = E_2(\vec{k})$,
the Chern number changes and that this point is 
an exeptional point where both eigenvectors point in the same direction, see 
Suppl.~Mat.\ \cite{supplement}.
Calculating the Chern numbers, we find that $C_\text{top} = C_\text{eff} = 1$ below the grey shaded area in 
Fig.\ \ref{fig:bool}(b), $C_\text{top} = 1$ and $C_\text{eff} = 0$ within the shaded area, and 
$C_\text{top} = C_\text{eff} = 0$ above. 

The two-body approach in the full Hilbert space encounters the
problem of a unique identification of the dressed single-boson states.
We attempted to identify them by (a)  maximizing the single-particle weight
or by (b) maximizing the overlap of adjacent eigenvectors as function of
wave vector in the BZ. But no reliable and numerical robust approach
could be established. Hence, we cannot check independently whether the
Chern number determined from the effective Hamiltonian possesses a 
physical meaning in the full Hilbert space. In addition, we investigated the
existence of localized edge modes for non-zero effective Chern number.
Both the IPR and the LDOS indicate that the potential edge modes delocalize
since they are not protected by an energy gap for overlapping bands, see 
Suppl.~Mat.\ \cite{supplement}.
Hence, bands overlapping with continua appear to lose their particular
topological properties, in line with the scenario
on the single-particle level \cite{malki19c,malki19b} in the absence of protecting
energy gaps.

\paragraph{\textbf{Conclusions}}

We have studied the topological properties of 
an interacting bosonic model on the honeycomb lattice 
allowing for the decay of a boson to two
bosons and the inverse fusion. While the ground state (vacuum)
is topologically trivial, the excited states are not. We investigated
how Chern numbers $C$ can be defined for these states and 
if edge states exist for non-vanishing $C$.

If the single-boson states
are renormalized by the continuum without overlap in energy, 
Chern numbers can be computed either 
in the many-body Hilbert space with the full Hamiltonian or in the 
single-boson space with the bilinear Hamiltonian plus the 
self-consistent self-energy (effective Hamiltonian).
It is the first key finding that
these two approaches agree while taking the self-energy at zero energy does not agree. 
This agreement of the effective self-consistent 
caculation with the correct Berry phases is corroborated by an analytic derivation.
Furthermore, our results indicate that the bulk-boundary correspondence holds in this case.

If both single-boson bands lie energetically
\emph{within} the two-boson continuum, one can extend the effective
determination to the non-Hermitian case. The Chern number changes at
gap closure which turns out to be an exceptional point.
In spite of intensive search, we could not establish a robust
definition of Berry phases in the full many-body Hilbert space.
This sheds doubts on the significance of non-trivial Chern numbers in case
of energetic overlaps. These doubts are enhanced by the absence of
edge modes indicated by our numerical results with vanishing IPR and
delocalized LDOS.

Our findings put the concept of non-trivial topology on a firm ground in 
presence of interactions inducing continua.
If the eigenstates are only renormalized by the hybridization with continua, Chern numbers 
and the bulk-boundary correspondence apply as in the non-interacting
case. If the single-particle states overlap with many-particle
continua the concept of Chern numbers can be extended to non-Hermitian effective Hamiltonians.
However, the Chern numbers are no longer protected by energy gaps and no localized edge modes seem to 
exist. This implies that an unambiguous definition of Berry phases in the full Hilbert space is no longer
possible.


B.H.~is grateful to the University of Manitoba and the Department of Physics and Astronomy 
for their hospitality and to the Studienstiftung des Deutschen Volkes for financial support.
J.S.~acknowledges support by the Natural Sciences and Engineering Research Council 
(NSERC, Canada) and by the Deutsche Forschungsgemeinschaft
(DFG) in Research Unit FOR 2316. G.S.U~thanks the DFG for support in UH 90-14/1.

\bibliography{main_vrs_2.bib}

\newpage 

{\Large\bf Supplemental Material}

\section{Self-Energy}
\label{a:self-energ}

Performing a Fourier transform and switching to the diagonal basis of the bilinear Hamiltonian $ {H}_0$, the Hamiltonian reads
\begin{align}
      H = &\sum_{\vec{k}, \alpha} E_\alpha(\vec{k}) \tilde{b}_{\vec{k},\alpha}^\dagger \tilde{b}_{\vec{k},\alpha} 
		\\ \nonumber
    &+ \sum_{\vec{k}, \vec{q}}\sum_{\alpha, \beta, \gamma} 
		\left( \tilde{g}_{\alpha \beta \gamma} (\vec{k}, \vec{q}) \tilde{b}_{\vec{k} 
		+ \vec{q}, \alpha}^\dagger \tilde{b}_{-\vec{q}, \beta}^\dagger \tilde{b}_{\vec{k}, \gamma} + 
		\text{h.c.} \right).
\end{align}
Here, $ b_{\vec{k}, \alpha} = \sum_\beta U(\vec{k})_{\alpha, \beta} \tilde{b}_{\vec{k}, \beta}$ where 
the unitary matrix $U(\vec{k})$ is chosen such that the bilinear part is diagonal with energies 
$E_\alpha(\vec{k}) = d_0(\vec{k}) + (-1)^\alpha d(\vec{k})$ for $\alpha \in \{1,2\}$, 
$d(\vec{k}) = \sqrt{d_x(\vec{k})^2 + d_y(\vec{k})^2 + d_z(\vec{k})^2}$ and 
\begin{subequations}
\begin{align}
    &d_0(\vec{k}) = 2 t_2 \cos(\phi) \left( \cos(\vec{k} \cdot \vec{a}_1) + \cos(\vec{k} \cdot \vec{a}_2) 
		+ \cos(\vec{k} \cdot \vec{a}_3) \right) \nonumber \\ &\qquad \qquad + E_0, 
		\\ 
    &d_x(\vec{k}) = t_1 \left( 1 + \cos(\vec{k} \cdot \vec{a}_1) + \cos(\vec{k} \cdot \vec{a}_2) \right), \\
    &d_y(\vec{k}) = t_1  \left( \sin(\vec{k} \cdot \vec{a}_1) + \sin(\vec{k} \cdot \vec{a}_2) \right), \\
    &d_z(\vec{k}) = M + 2 t_2 \sin(\phi) \big(\sin(\vec{k} \cdot \vec{a}_1) - \sin(\vec{k} \cdot \vec{a}_2) 
		\nonumber\\ & \qquad \qquad- \sin(\vec{k} \cdot \vec{a}_3) \big),
\end{align}
\end{subequations}
where we chose the primitive vectors $\vec{a}_1 = \frac{1}{2} (3, \sqrt{3})^T$ and $\vec{a}_2 = \frac{1}{2} (3, -\sqrt{3})^T$. The coupling constants in this basis are given by  
\begin{align}
    \tilde{g}_{\alpha \beta \gamma}(\vec{k}, \vec{q}) &=\frac{g}{\sqrt{N}} \big( (U(\vec{k} + \vec{q})_{1, \alpha})^* (U(-\vec{q})_{1, \beta})^* U(\vec{k})_{1, \gamma} \nonumber \\ 
		&+ (U(\vec{k} + \vec{q})_{2, \alpha})^* (U(-\vec{q})_{2, \beta})^* U(\vec{k})_{2, \gamma}\big).
\end{align}

At $T=0$, the spectral representation of the retarded Green's function reads
\begin{equation}
    G_{\alpha \beta}(\vec{k},\omega) = \langle 0 | b_{\vec{k},\alpha}  \frac{1}{\omega + i0^+ -  {H} + \epsilon_0} b_{\vec{k},\beta}^\dagger | 0 \rangle, \label{eq3:resolgf}
\end{equation}
where $\epsilon_0$ is the energy of the ground state $|0\rangle$; it is zero in our case. In the diagonal 
basis of 
$ {H}_0$ for fixed $\vec{k}$ the full Hamiltonian $ {H}$ on $ \mathcal{H}(1) 
\oplus \mathcal{H}(2)$ reads
\begin{equation} 
\label{eq3:2bodyhamil}
    H(\vec{k}) = \begin{pmatrix}
        \begin{matrix}
        D(\vec{k})
        \end{matrix}
        & \rvline & V^\dagger(\vec{k})   \\
      \hline
        V(\vec{k}) & \rvline &
        \begin{matrix}
        A(\vec{k})
        \end{matrix}
      \end{pmatrix}.
\end{equation}
Then, the self-energy can be exactly determined to be
\begin{equation}
    \Sigma(\vec{k}, \omega) = V(\vec{k})^\dagger \left(\omega - D(\vec{k})\right)^{-1} V(\vec{k}).
\end{equation}
In components, 
\begin{subequations}
\begin{align}
    &\Sigma(\vec{k}, \omega)_{\gamma, \delta} =
		\nonumber \\
		&\sum_{\vec{q}}\sum_{\alpha, \beta} C (\vec{k}, \vec{q}) \frac{\tilde{g}_{\alpha \beta \gamma}(\vec{k},\vec{q})^* \tilde{g}_{\alpha \beta \delta}(\vec{k},\vec{q})}{\omega + i0^+ - E_\alpha(\vec{k} + \vec{q}) -  E_\beta(-\vec{q})},\\
   & \text{with} \qquad C (\vec{k}, \vec{q}) = \begin{cases*}
        2 & if  $\vec{k} + 2 \vec{q} \in \bar{\mathcal{B}}$,  \\
               1 & else.
     \end{cases*}, \label{eqApp1:selfenergy}
\end{align}
\end{subequations}
where $\bar{\mathcal{B}}$ denotes the reciprocal lattice. 

\begin{figure}[htb]
    \centering
		\includegraphics[width=\columnwidth]{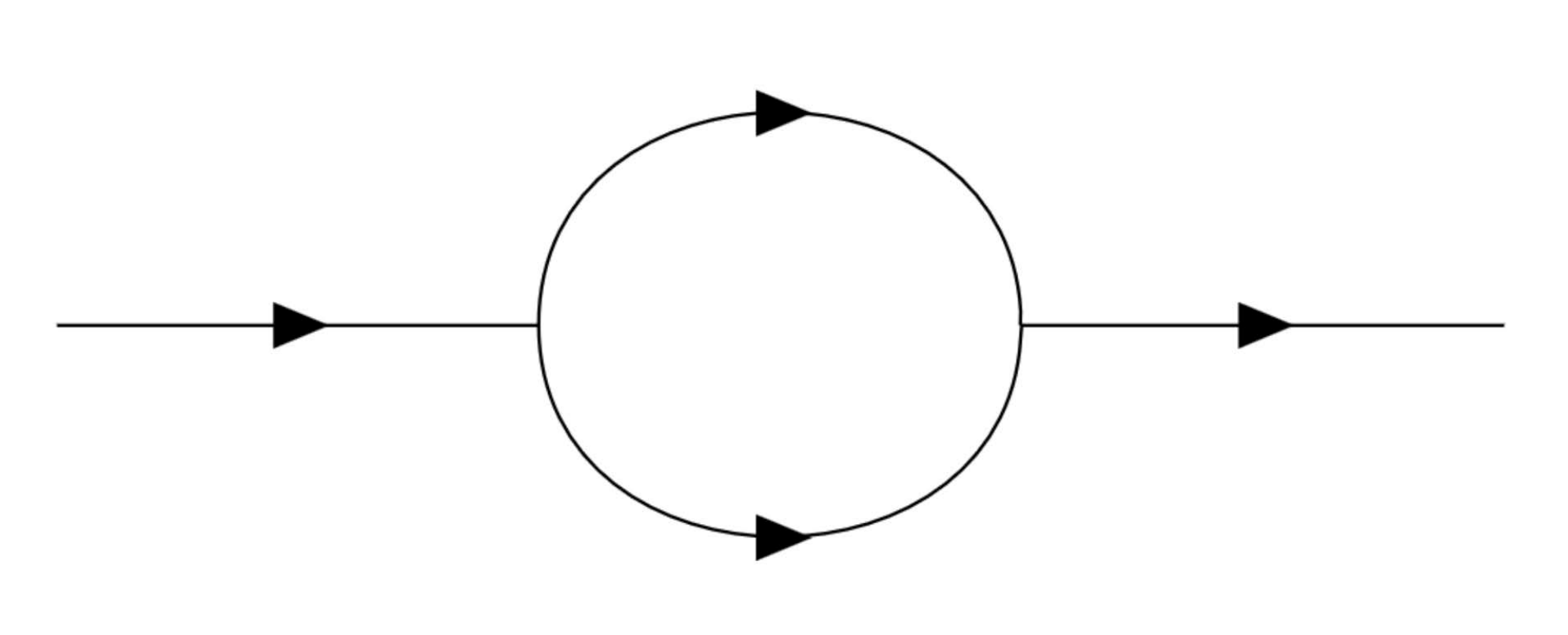}
    \caption{The only diagrammatic contribution for the case $\mathcal{H} = \mathcal{H}(0) \oplus \mathcal{H}(1) \oplus \mathcal{H}(2)$, 
		which is of second order in $g$. Neither higher-order diagrams nor vertex corrections
		are contributing for the considered restricted Hilbert space.}
    \label{figApp1:diag}
\end{figure}

This result can also be understood in perturbation theory at zero temperature. The only contributing diagram is shown in Fig.\ \ref{figApp1:diag}. Higher-order diagrams do not contribute since they would involve three or more propagators 
between two times which vanish due to the restriction of the Hilbert space to
at maximum two bosons. 
Also, no vertex corrections occur.

\section{Generalized Lanczos Tridiagonalization and Continued Fraction Representation}
\label{a:gen-lanczos}

The numerical evaluation of the sum over the free momentum in Eq.\ \eqref{eqApp1:selfenergy} depends on the discretization 
of the Brillouin zone. Generic results stay very spiky and do not allow for an interpretation based on a smooth
imaginary part unless the  $\delta$-peaks are artificially broadened by a finite imaginary part of the frequency
argument. To circumvent these issues, we generalize the continued fraction representation of scalar
self-energies and propagators \cite{petti85,viswa94} to matrices. As in the scalar case we find that
an appropriate terminator leads to smooth multi-dimensional self-energies, essentially independent
from the discretization, allowing for further evaluations as done in the main text.

We consider
\begin{equation}
\Sigma = V^\dagger \left(\omega - H \right)^{-1}V
\end{equation}
where $V \in \mathbb{C}^{N \times d}$ and $H \in \mathbb{C}^{N\times N}$. Hence, the self-energy
is a $d\times d$ matrix. The well-established scalar case 
is recovered for $d=1$. For the calculations in the main text we need $d=2$, but the following
derivation is valid for general $d$.
The generalized Lanczos algorithm yields a block tridiagonal matrix with $d \times d$ blocks such that 
\begin{equation}
\label{eq:block-structure}
    H_T := U^\dagger H U = \left(\begin{array}{cccc}
        A_0 & B_1^\dagger & 0 & \cdots \\
        B_1 & A_1 & B_2^\dagger & \ldots \\
        0 & B_2 & A_2 & \ldots \\
        \vdots & \vdots & \vdots & \ddots
        \end{array}\right),
\end{equation}
with $A_i, B_i \in \mathbb{C}^{d\times d}$ and after $L$ iterations, $U$ is of the form 
\begin{equation}
    U = (V_0, V_1, ..., V_L) \in \mathbb{C}^{N \times dL}.
\end{equation}
The very starting point is defined by $V_0=V$ if $V^\dag V=\mathds{1}_d$. Otherwise, we express
$V=V_0B_0$ by means of a standard QR-decomposition \cite{stoer78} where $V_0$ takes the role of $Q$ and $B_0$ the role
of $R$. Then, $V_0^\dag V_0=\mathds{1}$ holds by construction.

We aim at a proof by induction over $n\in\mathbb{N}_0$. Hence, we assume that we have shown the following two relations up to and including $j,m\le n$
\begin{subequations}
\label{eq:induction}
\begin{align}
\label{eq:ortho}
V^\dag_j V_m &= \mathds{1}_d \delta_{j,m} & \forall j,m \ge 0
\\ 
\label{eq:block}
V^\dag_j HV_m &= 0 & \forall |j-m|>1.
\end{align}
\end{subequations}

\paragraph{Base case} For $n=0$, $V_0$ has the desired property by construction and
the structure of the Hamiltonian \eqref{eq:block-structure} implies
\begin{equation}
A_0  = V_0^\dag H V_0 .
\end{equation}

\paragraph{Induction step} 
Assuming that \eqref{eq:induction} holds up to $n$,
\begin{subequations}
\begin{align}
\label{eq:A}
A_j &= V_j^\dag H V_j
\\
\label{eq:B}
B_j &= V^\dag_j H V_{j-1}
\end{align}
\end{subequations}
holds by construction. To pass to $n+1$ we consider
\begin{align}
\label{eq:S}
     S_n &= H V_n - V_n A_n - V_{n-1} B_{n}^\dag = V_{n+1} B_{n+1},
\end{align}
where the second equality refers to the $QR$-decomposition of $S_n$;
the diagonal elements of $B_{n+1}$ are chosen to be positive.
We exclude the possibility that one of these elements vanishes for the time being.
Note that we set $V_{-1} := 0$ to maintain the general validity of this step.

We show \eqref{eq:ortho} in three steps. First, we multiply \eqref{eq:S}
from the left with  $V_n^\dag$ yielding
\begin{subequations}
\begin{align}
\label{eq:vv1}
V_n^\dag V_{n+1} B_{n+1} &= V_n^\dag H V_n - V_n^\dag V_n A_n - V_n^\dag V_{n-1} B_{n}^\dag \\
 &=0 ,
\end{align}
\end{subequations}
where the first term on the right hand side of \eqref{eq:vv1}
cancels with the second using \eqref{eq:ortho} and \eqref{eq:A}
while the last one vanishes due to \eqref{eq:ortho}. Thus, we know 
\begin{equation} 
V_n^\dag V_{n+1} = 0=  V_{n+1}^\dag V_n.
\end{equation}
Second, we multiply \eqref{eq:S}
from the left with  $V_{n-1}^\dag$ yielding
\begin{subequations}
\begin{align}
\label{eq:vv2}
& V_{n-1}^\dag V_{n+1} B_{n+1} & \nonumber \\ 
&= V_{n-1}^\dag H V_n - V_{n-1}^\dag V_n A_n - V_{n-1}^\dag V_{n-1} B_{n}^\dag\\
 & =0,
\end{align}
\end{subequations}
where the first term on the right hand side of \eqref{eq:vv2} cancels  with the last term using \eqref{eq:ortho} and \eqref{eq:B}
while the second term vanishes due to \eqref{eq:ortho}. Thus, we know 
\begin{equation} 
V_{n-1}^\dag V_{n+1} = 0=  V_{n+1}^\dag V_{n-1}.
\end{equation}
Third, we multiply \eqref{eq:S}
from the left with  $V_m^\dag$ with $0\le m< n-1$ which results in
\begin{subequations}
\begin{align}
\label{eq:vv3}
& V_m^\dag V_{n+1} B_{n+1} \nonumber \\
&= V_m^\dag H V_n - V_m^\dag V_n A_n - V_m^\dag V_{n-1} B_{n}^\dag
\\ 
&= 0,
\end{align}
\end{subequations}
where the first term on the right hand side of \eqref{eq:vv3} vanishes due to \eqref{eq:block} and the 
second and third term vanish due to \eqref{eq:ortho}. Thus, we know 
\begin{equation} 
V_m^\dag V_{n+1} = 0=  V_{n+1}^\dag V_m.
\end{equation}
Hence, we know that \eqref{eq:ortho} holds also up to $n+1$.

What is left to complete the induction step is to prove $V_{n+1}^\dag HV_m=0$ for
$0\le m<n$.  To this end, we multiply \eqref{eq:S} for $n=m$ from the left with $V_{n+1}^\dag$
yielding
\begin{subequations}
\begin{align}
0 &= V_{n+1}^\dag V_{m+1} B_{m+1}\\
\label{eq:middle}
 &= V_{n+1}^\dag H V_m - V_{n+1}^\dag V_m A_m - V_{n+1}^\dag V_{m-1} B_{m}^\dag 
\\
 &= V_{n+1}^\dag H V_m.
\end{align}
\end{subequations}
The first identity ensues from \eqref{eq:ortho} which we just have shown to hold. The second and 
third terms in \eqref{eq:middle} vanish also due to \eqref{eq:ortho} so that the
vanishing of the blocks beyond the minor diagonals in the Hamiltonian matrix is also 
extended to $n+1$. Recall that $V_m^\dag H V_{n+1}=0$ follows from hermiticity.
This concludes the proof by induction.

Above, we assumed that the $B_n$ always display the full rank of $d$.
We point out that it can happen that for some $n=n_0$, one column of $V_n$
vanishes. Then, the above algorithm reduces to a $(d-1)-$dimensional Lanczos
algorithm for $n > n_0$.

\paragraph{Multidimensional continued fraction expansion}
Given a basis in which $H$ is block tridiagonal, we derive the continued fraction representation 
of the self-energy. We start for block matrices from the general identity 
\begin{equation}
    \left(\begin{array}{ll}
        {A} & {B} \\
        {C} & {D}
        \end{array}\right)^{-1} =\left(\begin{array}{cc}
        \left({A}-{B D}^{-1} {C}\right)^{-1} & * \\
        * & *
        \end{array}\right),
\end{equation}
where we assume the existence of the inverse on the left hand side and of $D^{-1}$; the blocks $*$ do not matter. Iterating this relation for the blocks of $H_T$ we obtain
\begin{subequations}
\begin{align}
    \Sigma &= V^\dagger \frac{1}{\omega - H} V \\
           &= B_0^\dagger P^\dag \left( \frac{1}{\omega - H_T} \right) P B_0\\
           &= B_0^\dagger \left(\omega - A_0 - B_1^\dagger 
					\left(\omega - A_1 -\ldots\right)^{-1} B_1 \right)^{-1} B_0,
\end{align}
\end{subequations}					
where  we have used $V=V_0B_0$ and $P=(\mathds{1}_d, 0)^\dag \in \mathbb{C}^{N\times d}$. 
This relation can be expressed best as recursion
\begin{align}
    \Sigma_n &= B_n^\dagger \frac{1}{\omega - A_n - \Sigma_{n+1}} B_n
\end{align}
with $\Sigma=\Sigma_0$. This provides the continued fraction expansion
generalized to matrices.

\paragraph{Termination}
Finally, we discuss the issue of how to terminate the continued fraction expansion.
Stopping at a finite depth of the fraction $n_0$, i.e., setting $\Sigma_{n_0}=0$
yields an imaginary part of the self-energy made up from a finite set of $\delta$-distributions. For
$d=1$ there are $n_0$ poles and generally we expect $d n_0$ poles. But this is still
unsatisfactory because no smooth function is generated. In the scalar case of $d=1$
the square-root terminator, see Refs.\ \cite{petti85,viswa94} as well as below in Eq.\
\eqref{eq:terminator},
allows one to terminate the continued fraction at finite depth while still 
yielding a smooth function. Here we briefly discuss whether an equivalent
procedure is also possible in the multidimensional case. The answer is a
yes, but less generally.

We assume that the following limits exist
\begin{equation}
    \lim_{n \to \infty} A_n =: A_\infty, \qquad \lim_{n \to \infty} B_n =: B_\infty
\end{equation}
in the sense that each matrix element converges to a complex number. In order to derive a terminator, 
the following equation has to be solved
\begin{equation}
    \Sigma_\infty = B_\infty^\dagger \frac{1}{\omega - A_\infty - \Sigma_{\infty}} B_\infty,
\end{equation}
which is equivalent to
\begin{equation}
\label{eq:quad-matrix}
    \Sigma_\infty {B_\infty^{-1}}^\dagger \Sigma_\infty - \left( \omega - A_\infty \right){B_\infty^{-1}}^\dagger \Sigma_\infty + B_\infty = 0.
\end{equation}
We multiply ${B_\infty^{-1}}^\dagger$ from the left and define 
\begin{subequations}
\begin{align}
    X &:= {B_\infty^{-1}}^\dagger \Sigma_\infty,
		\\
		O &= -{B_\infty^{-1}}^\dagger \left( \omega - A_\infty \right) {B_\infty^{-1}}^\dagger, 
		\\ P &:= {B_\infty^{-1}}^\dagger B_\infty,
\end{align} 
\end{subequations}
to obtain the quadratic matrix equation 
\begin{equation} 
\label{matrixeq}
    X^2 + O X  + P = 0.
\end{equation}
As discussed in Ref. \cite{higha00}, the solution of such an equation is not straightforward. 
For this reason, no general terminating procedure can be provided.

In our calculations, we observed, however, that often the block matrices $A_n$ and $B_n$ become
more and more diagonal as $n$ increases so that 
\begin{equation}
    (A_\infty)_{ij} = (B_\infty)_{ij} \approx 0 \quad \text{if} \quad i \neq j.
\end{equation}
Then, a simple solution suggests itself. The approximate diagonality of the Lanczos coefficients imply that 
$(\Sigma_\infty)_{ij} \approx 0$ so that Eq.\ \ref{matrixeq} reduces to $d$ independent equations for the diagonals, each solved by the standard square-root terminator 
\begin{equation}
\label{eq:terminator}
    \Sigma_{\infty} \!= \! \begin{cases}\frac{\omega-a_{\infty}}{2}
		-\sqrt{\left(\frac{\omega-a_{\infty}}{2}\right)^2-b_{\infty}^2}, 
		&\omega>E^{(2)}_\text{max} , \\ \frac{\omega-a_{\infty}}{2}-\mathrm{i} 
		\sqrt{b_{\infty}^2-\left(\frac{\omega-a_{\infty}}{2}\right)^2}, 
		&\omega \in\left[E^{(2)}_\text{min}, E^{(2)}_\text{max} \right], 
		\\ \frac{\omega-a_{\infty}}{2}+\sqrt{\left(\frac{\omega-a_{\infty}}{2}\right)^2-b_{\infty}^2}, 
		&\omega<E^{(2)}_\text{min} ,\end{cases}
\end{equation}
for real values of $\omega$
where $E^{(2)}_\text{min/max}$ corresponds to the minimum/maximum of the spectrum of $H$
and $a_\infty$ and $b_\infty$ are the diagonal element of $A_\infty$ and $B_\infty$, respectively.
They are given by \cite{petti85,viswa94}
\begin{equation}
 \label{eq:terminator_ab}
a_\infty = (E^{(2)}_\text{max}+E^{(2)}_\text{min})/2 \qquad
b_\infty = (E^{(2)}_\text{max}-E^{(2)}_\text{min})/4.
\end{equation}
For complex frequencies $z\in\mathbb{C}^+$, i.e., $\Im z >0$, $\Sigma_\infty$ 
is given by the retarded solution of the quadratic equation
\begin{equation}
\label{eq:terminator_C}
 \Sigma_\infty^2 -(z-a_\infty)\Sigma_\infty + b_\infty^2 = 0,
\end{equation}
i.e., the one with negative imaginary part. This is the scalar version, i.e., for $d=1$, 
of Eq.~\eqref{eq:quad-matrix}.
In this work, we used this termination procedure as an approximation for the general matrix case
yielding robust and smooth results.

\section{Equivalence of the full eigenvalue problem and 
the self-consistent diagonalization of $H_\text{eff}$}

We consider the Hilbert space consisting of the direct sum of the 
single-particle space $\mathcal{H}(1)$ and the remainder $\mathcal{H}\rs$. The Hamiltonian
reads
\begin{equation}
H =H_1+ H\rs + H_V,
\end{equation}
where $H_1$ only acts in $\mathcal{H}(1)$, $H\rs$ only in $\mathcal{H}\rs$ while
$H_V$ links both according to $P\rs H_V P_1=V$ where $P_1$ projects onto $\mathcal{H}(1)$
and $P\rs$ projects onto $\mathcal{H}\rs$.

The full eigenvalue problem reads
\begin{equation}
\label{eq:full-eigen}
H(v_1+v\rs) = E (v_1+v\rs)
\end{equation}
where the vectors $v_1$ and $v\rs$ live in  $\mathcal{H}(1)$ and in $\mathcal{H}\rs$, respectively.
Hence, \eqref{eq:full-eigen} projected onto $\mathcal{H}(1)$ and onto $\mathcal{H}\rs$
yields
\begin{subequations}
\begin{align}
\label{eq:1}
H_1v_1+V^\dag v\rs &= Ev_1
\\
H\rs v\rs+V v_1 &= Ev\rs.
\label{eq:2}
\end{align}
\end{subequations}
The second equation is equivalent to $(E-H\rs)^{-1} V  v_1 = v\rs$ which is
well defined in Case I because we consider energies $E$ in the range of the single-particle energies
well below the lower continuum edge given by the lowest eigenvalue of $H\rs$.
Inserting $v\rs$ in the first equation \eqref{eq:1} yields 
\begin{equation}
\label{eq:self-eigen}
(H_1 + \Sigma(E)) v_1 = E v_1,
\end{equation}
where we use the self-energy $\Sigma(E) =V^\dag (E-H\rs)^{-1} V$. Clearly,
the solution of \eqref{eq:self-eigen} requires to find the self-consistency
for $E$. Its solution is equivalent to solving the
full eigenvalue problem \eqref{eq:full-eigen}.

\section{Edge States on a Honeycomb Ribbon}

\begin{figure}[htb]
    \centering
		\includegraphics[width=\columnwidth]{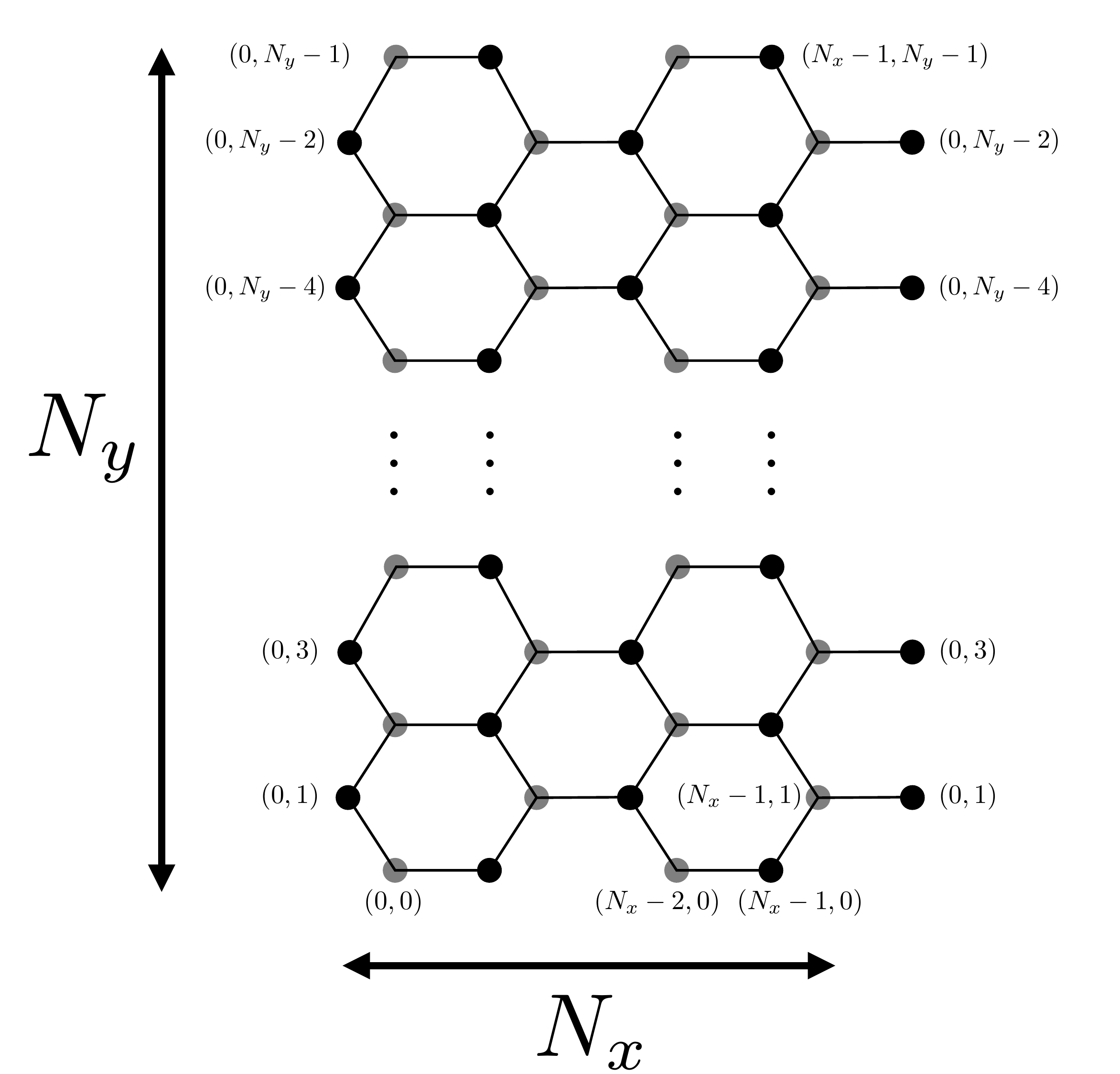}
    \caption{Sketch of the considered lattice which is periodic in $x$ direction (horizontal) 
		with length $N_x$ and has open boundary conditions and a finite length $N_y$ in 
		$y$ direction (vertical). The total number of sites is $N_xN_y$ where $N_x$ is even 
		due to the basis of two sites and $N_y$ is odd. The pair $(i, j)$ denotes the
position of the site with $0 \le  i < N_x$ and $ 0\le j < N_y$. 
}
    \label{fig:ribbon}
\end{figure}

\begin{figure}[htb]
    \centering
			 \includegraphics*[width=\columnwidth]{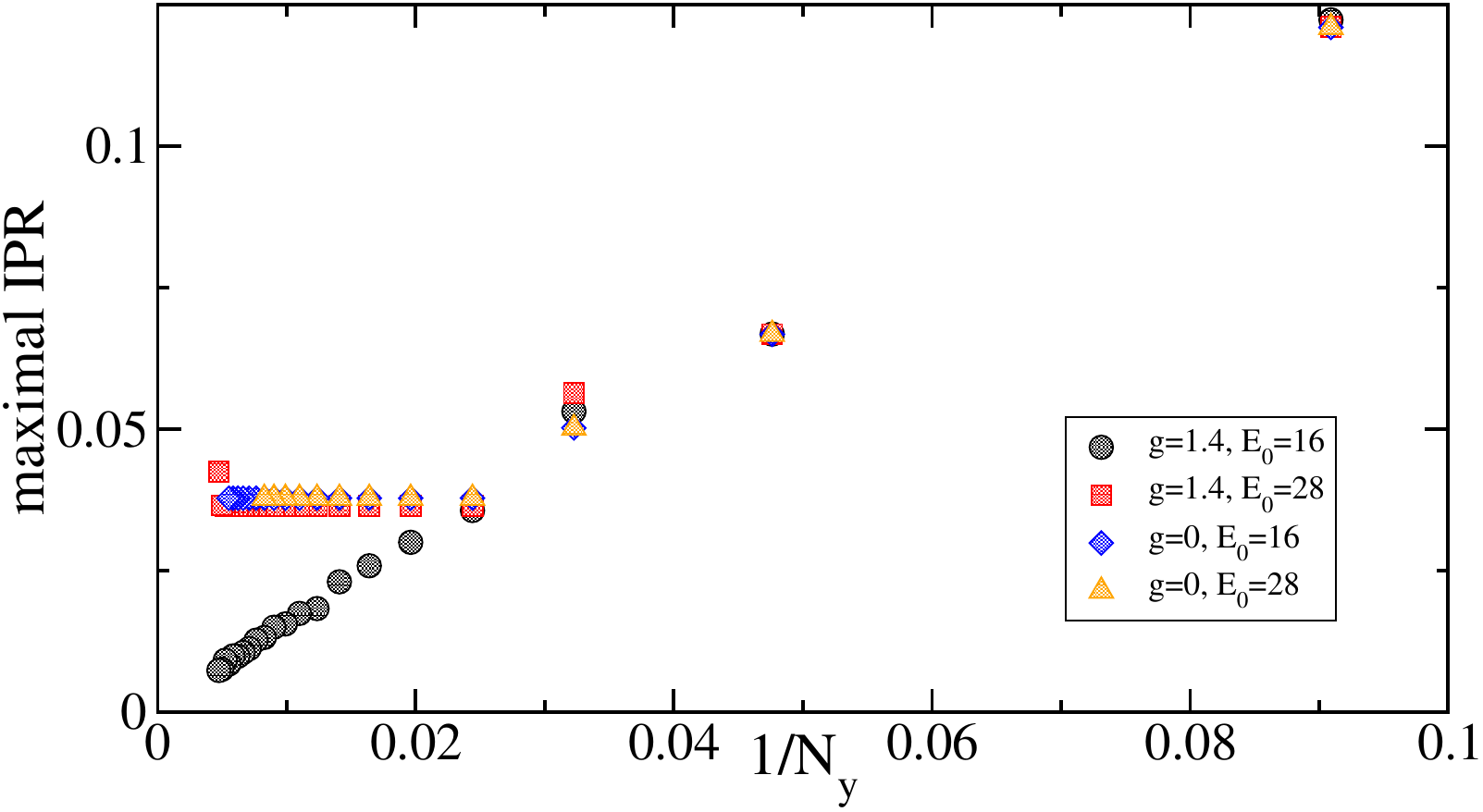} 
     \caption{IPR for a ribbon with $N_x=2$ 
		for vanishing and for finite interactions at parameters given in the legend.
		The states are localized without interaction and also in the interacting case for large $E_0=28$ avoiding energetic
		overlap. For small $E_0=16$ the overlap with the continuum leads to delocalization.}
    \label{fig:ipr}
\end{figure}

To answer the question whether non-trivial Chern numbers imply edge states in the interacting case 
similar to the non-interacting case, we consider the model defined in Eq.~\eqref{eq:modeldef} 
on the honeycomb ribbon with $N_xN_y$ sites as sketched in Fig.~\ref{fig:ribbon}. 
One signature of localization is a finite inverse participation ratio $I$ of eigenstate $n$ 
defined by
\begin{equation}
\label{def:IPR}
    I_n = \frac{\sum'_i |\psi^{(1)}_n(x_i,y_i)|^4}{\left(\sum'_i |\psi^{(1)}_n(x_i,y_i)|^2\right)^2},
\end{equation}
where $\psi^{(1)}_n(x_i,y_i)$ is the wave function of the $n$-th eigenstate projected onto $\mathcal{H}(1)$ in position space. Each site $i$ has the spatial coordinates $(x_i,y_i)$.
The sum $\sum'$ runs over all $y_i$ and the two $x_i$ in the unit cell of the ribbon.  
From this definition, it is easy to see that a totally local
state where only one site has the finite amplitude 1 yields $I_n=1$ while a  state completely
delocalized over $N$ states implies $I_n=1/N$. 
In Fig.\ \ref{fig:ipr}, we show the significantly different behavior of the IPR of a localized
state and of a delocalized state.

In addition, we computed the local density of states (LDOS).
The LDOS at zero temperature reads 
\begin{equation}
    A(x_i, y_i;\omega) = \sum_n |\langle n | b_i^\dagger | 0 \rangle|^2 \delta(\omega - E_n),
\end{equation}
where $ {H} |n \rangle = E_n |n \rangle$.
Since we are interested in the localization at the edges $y = 0$ and $y = N_y-1$, 
where $N_y$ is the finite width
 of the ribbon in $y$ direction, we additionally sum over the $x_i$ coordinates, yielding 
\begin{equation}
    A(y_i;\omega) = \frac{1}{N} 
		\sum_n \sum_{x_i} |\langle n | b_i^\dagger | 0 \rangle |^2\delta(\omega - E_n),
\end{equation}
which is the quantity of interest in the following.

\paragraph{Non-interacting case}

\begin{figure}[htb]
    \centering
		\includegraphics[width=\columnwidth]{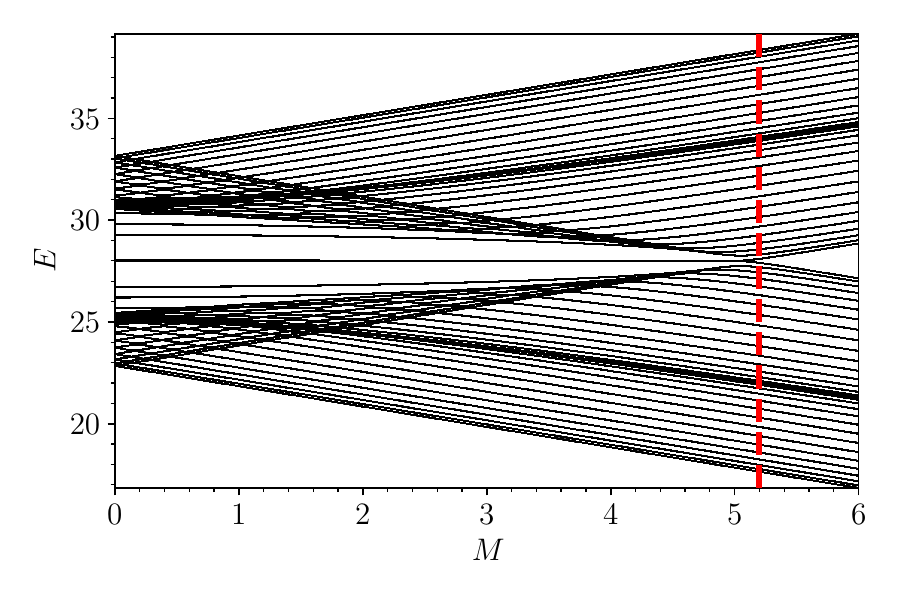}
    \caption{Spectrum of the non-interacting model on a honeycomb ribbon with $N_x=2, N_y = 33$. 
		The  vertical red line is the topological phase boundary at $M = 3\sqrt{3}$.}
    \label{fig:specnonint}
\end{figure}

For completeness and comparison, we first discuss the well-known 
non-interacting case of Eq.\ \ref{eq:modeldef}. We choose $N_x=2$ and 
\begin{equation}
    t_1 = t_2 = 1, \phi = \frac{\pi}{2}, E_0 = 28.
\end{equation}
As mentioned in the main text, the Chern number changes in this case 
from $1$ to $0$ at $M = 3 \sqrt{3}$ upon increasing $M$. 
We show the spectrum of $  H$  as function of $M$ in Fig.\ \ref{fig:specnonint}.

\begin{figure}[htb]
    \centering
		\includegraphics[width=\columnwidth]{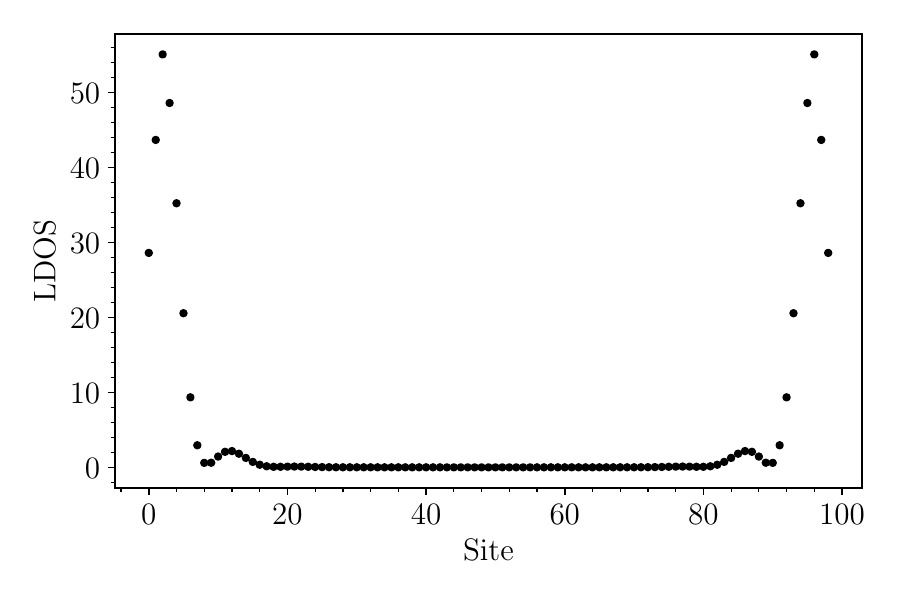}
    \caption{LDOS of one of the two  states with $E \approx E_0$
		in the topological phase of the non-interacting model 
		with $E \approx E_0$ and $M = 4.5$ for $N_x=2$ and $N_y=99$. 
		Clearly, the state is localized at the edges of the system,
		illustrating the bulk-boundary correspondence.}
    \label{fig:ldos_nonint}
\end{figure}

The spectrum is symmetric around $E_0$ and we find two states with $E \approx E_0$, 
which are the well-known edge  modes. Note that the two energies do not coincide 
perfectly because for any ribbon of finite width each state
localized at one edge overlaps weakly with the state localized at the other edge. The true
eigenstates are the symmetric and antisymmetric combination displaying a tiny, exponentially small energy splitting. The corresponding LDOS is shown in Fig.~\ref{fig:ldos_nonint} for $M = 4.5$ 
implying $C=1$. For $M > 3\sqrt{3}$ and thus $C=0$, we find that the states lying energetically the
closest to $E_0$ are not localized anymore and have their highest weight in the bulk (not shown).

\paragraph{Interacting case}

\begin{figure}[htb]
    \centering
		\includegraphics[width=\columnwidth]{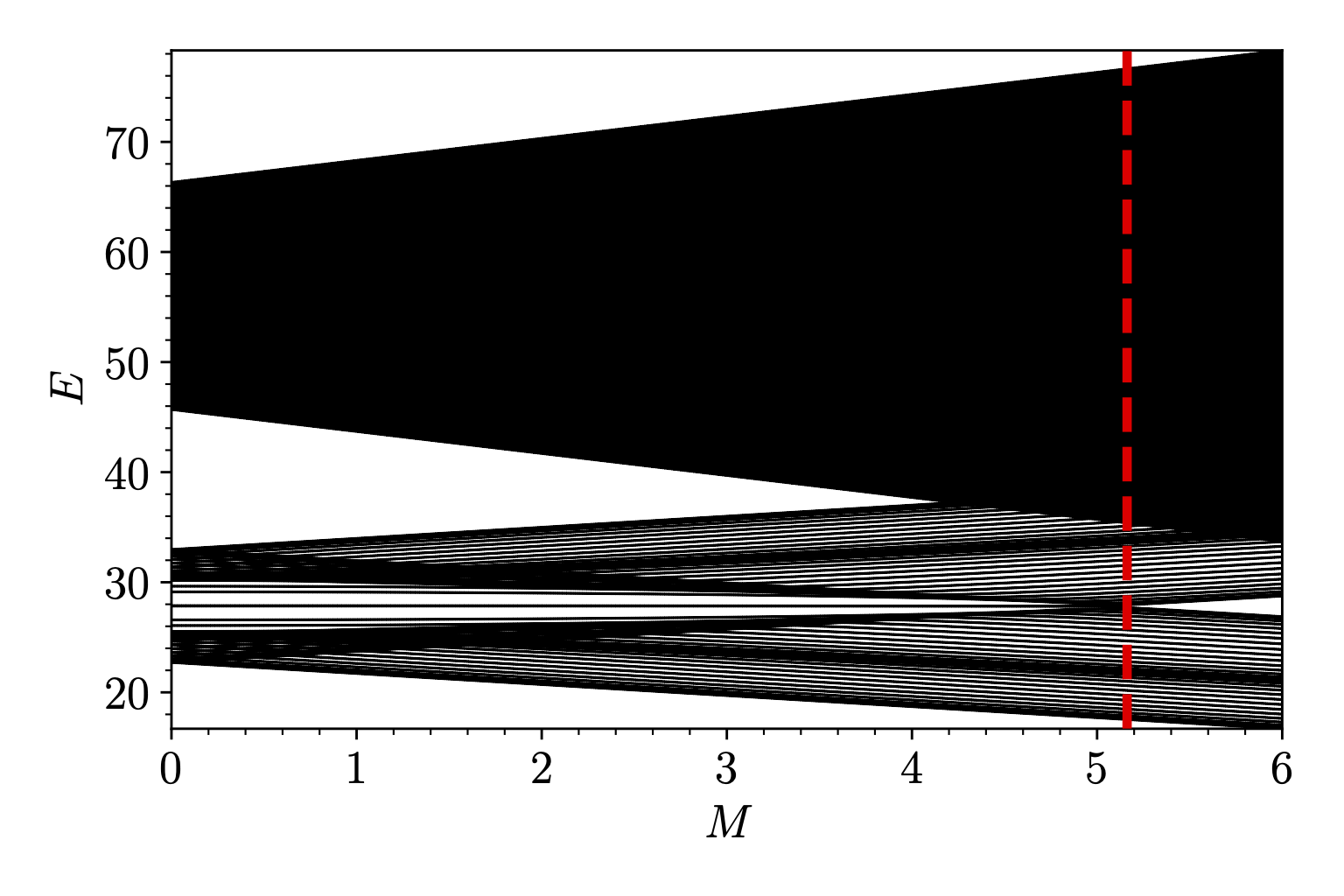}
    \caption{Spectrum of the interacting system with $E=28$ and $g=1.4$, i.e., for Case I,
		with $N_x=2$ and $N_y=33$. The vertical red line is the topological phase boundary 
		at $M = 5.162$, cf.\ Fig.\ 1.}
    \label{fig:specintE028}
\end{figure}

We turn on interactions and investigate the same interaction strength 
as considered in the main text, i.e., $g = 1.4$. We stress again, that $g \ll E_0$ and 
hence the interaction effects are still small compared to the bare energies, i.e.,
the system is in the perturbative regime.
In Fig.\ \ref{fig:specintE028}, we show the spectrum as a 
function of $M$ for Case I of the main text, realized by choosing $E_0 = 28$.

\begin{figure}[htb]
    \centering
	\includegraphics[width=\columnwidth]{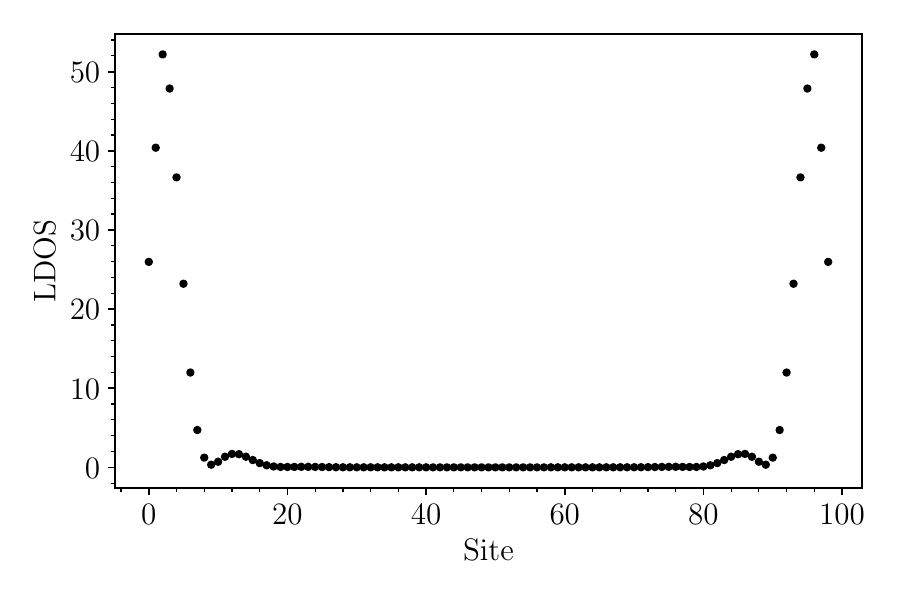}
   \caption{LDOS of an edge state for the interacting case shown in Fig.~1
	with $E \approx E_0 = 28$, i.e., 	Case I of the main text, 	and $N_x=2, N_y = 99$ sites. 
	The existence of edge states demonstrates that the 
	bulk-boundary correspondence still applies in  Case I.}
    \label{fig:ldos_int_E028}
\end{figure}


We observe two important points:
\begin{itemize}
    \item[1)] Two states with energy $E \approx E_0$ still exist within the interacting spectrum. 
    \item[2)] The gap between the states above and below the two states with $E \approx E_0$ 
		closes for a smaller value  $M_C$	than in the non-interacting case, 
		which is consistent with Fig.~1. 
\end{itemize}
Obvious candidates for edge states are again those with $E \approx E_0$.
The LDOS of one of these states is displayed in Fig.~\ref{fig:ldos_int_E028} 
for $M=4.5$ corresponding to Case I.
We conclude that in Case I where the two-particle continuum does not overlap with the 
single-particle bands of the bulk model, non-vanishing Chern numbers signal the existence 
of edge states as the bulk-boundary correspondence implies.

\begin{figure}[htb]
    \centering
		\includegraphics[width=\columnwidth]{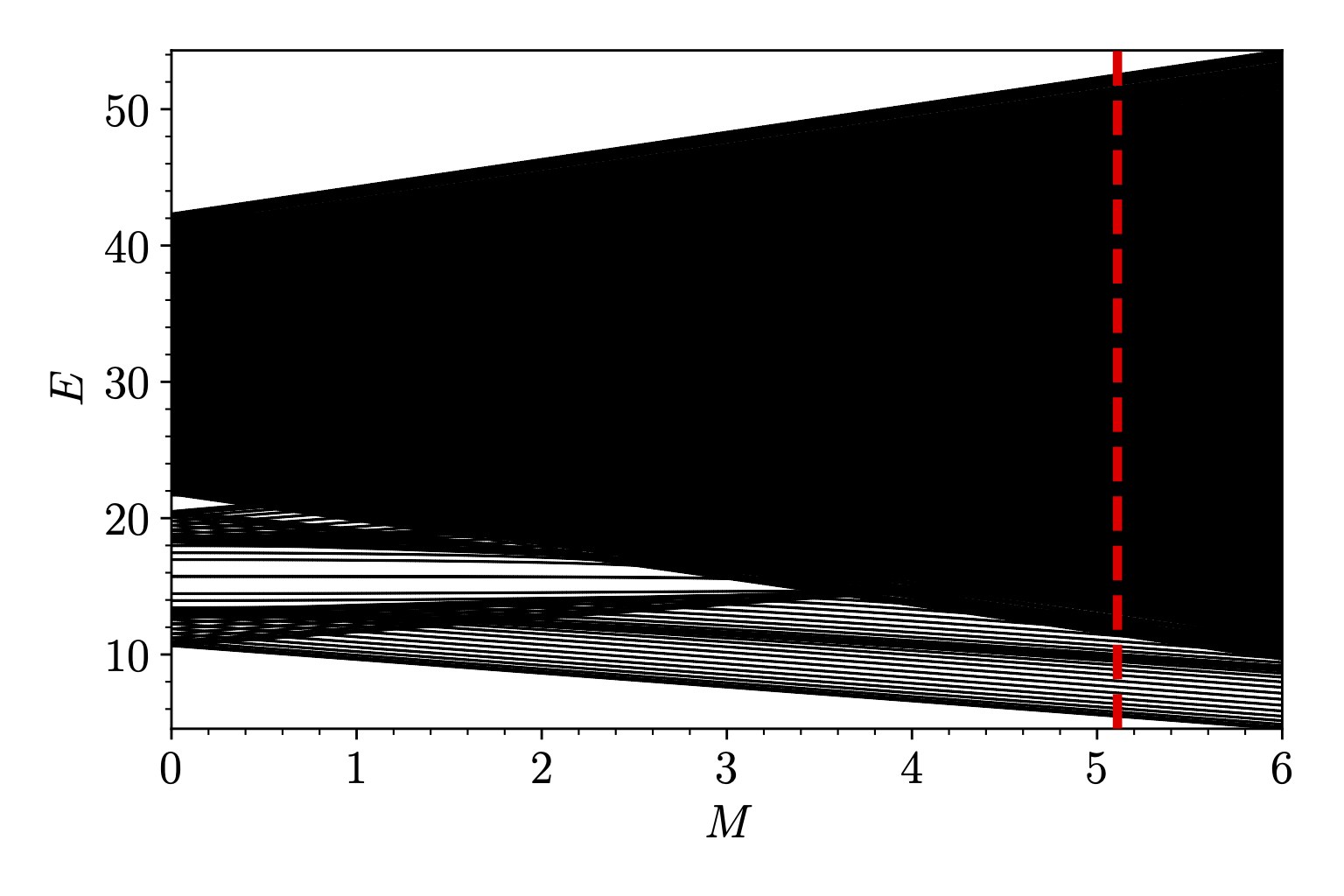}
    \caption{Spectrum of the interacting system with $E_0=16$ and $g=1.4$, i.e., for Case II,
		with $N_x=2$ and $N_y=33$. The vertical red line is the topological phase boundary 
		at $M = 5.110$ determined from the self-consistent solution of the effective 
		Hamiltonian.}
    \label{fig:specintE016}
\end{figure}

\begin{figure}[htb]
    \centering
		\includegraphics[width=\columnwidth]{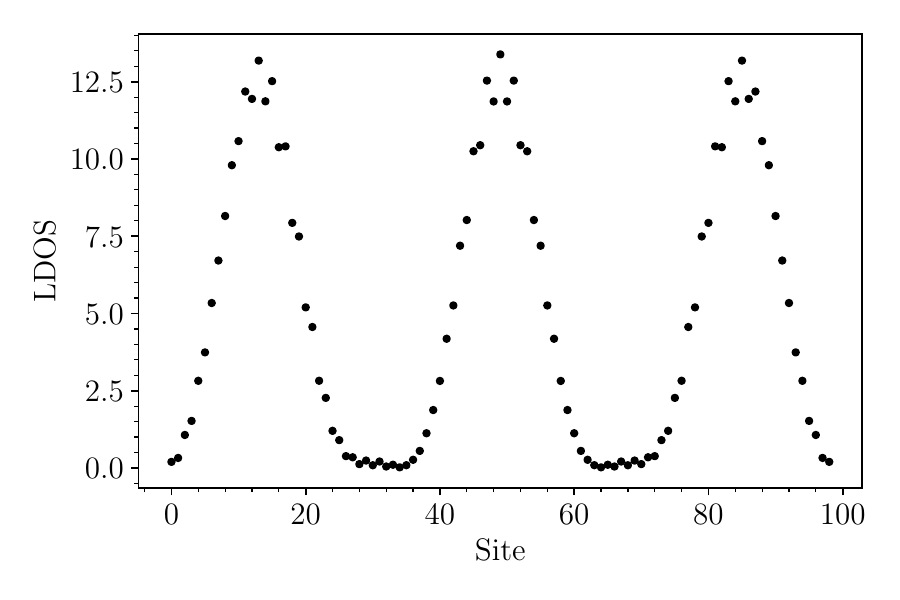}
    \caption{LDOS of the state with the maximum IPR given that the
		spectral weight of the one-boson sector represents more than $50\%$ 
		of the total spectral weightfor $N_x = 2$ and $N_y = 99$ with $M = 4.5$, i.e., 	
		for Case II.}
    \label{LDOS_maxIPR}
\end{figure}

Finally, we turn to Case II, where both bands overlap with the two-boson continuum. 
Note that we need $M > 3$ in order for the overlap to occur. 
The spectrum for this case is shown in 
Fig.~\ref{fig:specintE016}. For $M > 3$, states with $E \approx E_0$ hybridize with the two-particle continuum and candidates for edge states cannot be determined straightforwardly. 
In this case, we consider the state with the maximum IPR among those states which have
at least half their weight in the one-boson sector. In practice, we find that there is a clear distinction between predominantly single-particle states, where the single-particle weight is larger than $98\%$, and the states which are predominantly of two-particle character. Some of the latter states have tiny 
one-particle contributions which appear to be localized. However, their LDOS is so minute that 
they do not influence the physics in any way.
We study the scaling of the largest IPR among the preselected states
with increasing width of the ribbon.

We choose $t_1=t_2=1$, $\phi=\pi/2$, $E_0=16$, $g=1.4$, and $M=4.5$. 
This puts the system in the region where the analysis of the effective Hamiltonian
 indicates a non-zero Chern number while the single-particle bands are already hybridized 
with the continuum. The scaling of the largest IPR as a function of the length $N_y$ is shown in 
 Fig.~\ref{fig:ipr} (black circles) indicating delocalization
since this IPR clearly tends to zero for $N_y\to\infty$.
This is strongly corroborated by Fig.~\ref{LDOS_maxIPR} 
showing the LDOS of the state with maximal IPR.
Obviously, this LDOS belongs to a delocalized standing wave as one might have 
expected due to the open boundary conditions.
We conclude that the numerical data provide strong evidence that edge states 
delocalize when hybridizing with a continuum at the same energy.

\section{Topological phase transition at an exceptional point}

\begin{figure}[htb]
    \centering
 		\includegraphics[width=\columnwidth]{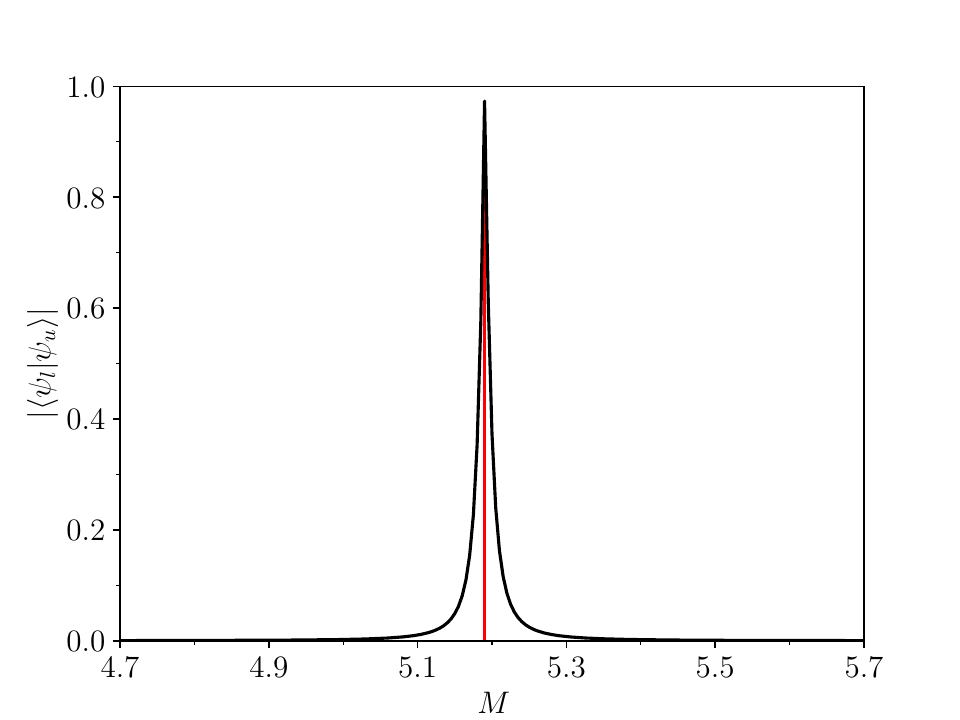}
    \caption{Absolute value of the overlap given by the scalar product between the 
		two eigenstates $|\psi_u\rangle$ and $|\psi_l\rangle$ corresponding to 
		the eigenvalues $E_u \ge E_l$ at $g=0.4$ and $E_0=16$ 
		as function of the alternating potential $M$. 
		The overlap is practically zero except for a narrow region around gap closure, marked
		by the vertical red line, which is accompanied by the change of the Chern number.}
    \label{fig:ep}
\end{figure}

In Case II, i.e., for overlap of the single bands with the continuum, the
effective eigenvalue problem Eq.~(3b) is no longer a hermitian one. Thus, the question
arises which angle the two two-dimensional eigenvectors enclose.
We studied this angle or the corresponding overlap,
respectively, numerically across the closing of the gap at the $K$ point
in the BZ where the Chern number changes from 1 to 0 upon increasing $M$. 
The resulting data is depicted in Fig.~\ref{fig:ep}.

The two eigenvalues merge
at $M=5.19$. Their eigenvectors are essentially perpendicular to each other, i.e.,
the non-hermiticity is hardly felt except for the vicinity of the topological
phase transition. There the overlap quickly rises to (almost) unity which implies
that the eigenvectors become parallel. This is the smoking gun characteristics of an exceptional
point. Thus, our numerical data indicate that the topological transition in the 
case of overlapping bands and continua, i.e., in the non-Hermitian regime, occurs
at exceptional points. We emphasize, however, that the frequency dependence of
the effective Hamiltonian plays an important role and cannot be neglected.

\end{document}